\documentstyle[12pt,epsfig]{article}

\begin{document}
\newcommand{\EENG}{\rm e^+ e^-\rightarrow n\gamma~ (n\geq 2)}
\newcommand{\EEG}{\rm e^+ e^-\rightarrow \gamma\gamma}
\newcommand{\EEGG}{\rm e^+ e^-\rightarrow \gamma\gamma(\gamma)}
\newcommand{\EEGGG}{\rm e^+ e^-\rightarrow \gamma\gamma\gamma}
\newcommand{\EEEEG}{\rm e^+ e^-\rightarrow e^+e^-(\gamma)}
\newcommand{\LAMP}{ \Lambda_{+}}
\newcommand{\LAMM}{ \Lambda_{-}}
\newcommand{\LAMS}{ \Lambda_{6}}
\newcommand{\LAMPP}{ \Lambda_{++}}
\newcommand{\LAMMM}{ \Lambda_{--}}
\newcommand{\DSDW}{ \sigma(\theta)}
\newcommand{\MESTAR}{ m_{{\rm e^{\ast}}}}
\newcommand{\EELL}{\rm e^+ e^-\rightarrow l^{+}l^{-}}
\begin{center}
\Large {\bf Size of Fundamental Particles and Selfgravitating
Particlelike Structure with the de Sitter Core}              \\
\vspace{3mm}
\large I.~Dymnikova$^{1}$, A.~Hasan$^{2}$, and J.~Ulbricht$^{2}$  \\
\vspace{2mm}
\small
$^{1}$Pedagogical University of Olsztyn, PL-10-561 Olsztyn, Poland \\
$^{2}$Eidgen\"ossische Technische Hochschule, ETH Z\"urich,
CH-8093 Z\"urich, Switzerland                                     \\
\vspace{5mm}
\parbox{14cm}{
     The processes $ \EENG $ are studied to estimate the total
     and differential cross sections of these reactions using
     the L3 data collected during 1991--1998 at energies ($\sqrt{s}$)
     in the range 91--183 GeV. The lower limits obtained at 95\% CL are,
     on a contact interaction energy scale $ \Lambda > 1065 $ GeV,
     on the mass of an excited electron $ \MESTAR > 263 $ GeV.
     The upper and lower limits on the gravitational size of fundamental particles
     are estimated  using the model of selfgravitating particlelike structure
     with de Sitter core, which gives an estimate for self-coupling of the Higgs field
     $\lambda\leq{\pi/16}$ and for the mass of the Higgs scalar $m_H\leq{154}$ GeV.}
\vspace{5mm}
\end{center}
\normalsize
\noindent
     The processes $ \EENG $ are studied to estimate the total
     and differential cross sections of these reactions using
     the L3 data collected during 1991--1998 at energies ($\sqrt{s}$)
     in the range 91--183 GeV.
The observed rates and distributions are in agreement with the QED
predictions. This puts constraints on the existence of an excited
electron of mass $ m_{e^{*}} $ which may replace the virtual
electron in the QED process [1] and also on 
the model with deviation from QED arising from an
effective interaction with non-standard
$ e^{+} e^{-} \gamma $ couplings and
$ e^{+} e^{-} \gamma \gamma $ contact terms [1] .
In the first case, the L3 data deliver
$ \MESTAR > 263 $ GeV with
the QED cut-off parameters $ \LAMP > 262 $ GeV and
$ \LAMM > 245 $ GeV [1] and
in the second case, limit the geometrical
diameter of the interaction area to
$ \Lambda > 1065 $ GeV (  $ 1.9\times10^{-17} $ cm ).                  \\
The L3 analysis [1] and the CDF [2] data
exclude excited electrons below $ 263 $ GeV
and excited quarks between $80$ and
$ 570 $ GeV and between $580$ and $760$ GeV.
The limits for the direct contact term measurements of
ref. [1] and the g-2 experiments
[3] are in the range $ 10^{-17} $cm - $ 10^{-22} $ cm.
As in the QED it seems that the fundamental particles (FPs)
have no internal substructure and their size is down to zero. \par
\noindent
To find some limits on a size of a FP we need a model for extended
particle.
In recent years a variety of models of selfgravitating structures
with non-Abelian fields has been found including black holes with
non-Abelian hair [4]. Among them there exists
a neutral type for which a
non-Abelian structure can be approximated by a sphere of
the uniform vacuum density $\rho_{vac}$ whose radius is the Compton
wave length of a massive non-Abelian field; numerical results suggest
that an additional horizon must exist and a black hole is approximated
near it by de Sitter-Schwarzschild spacetime [4]. \par
\noindent
De Sitter-Schwarzschild spacetime has been studied in the literature with
the original motivation
to replace a black hole
singularity by de Sitter regular core. The
exact analytic solution describing de Sitter-Schwarzschild spacetime was found
by one of us [5], and
it appeared to present three types of objects
dependently on a mass:
a nonsingular neutral black hole
with two horizons for mass $m>
m_{cr}\simeq{0.3m_{Pl}({\rho}_{Pl}/{\rho}_{vac})^{1/2}}$,
extreme black hole with
the degenerate horizon for $m=m_{cr}$,
and neutral particlelike structure
without horizons for $m<m_{cr}$ (see Fig.1).
In the course of Hawking evaporation
a black hole loses its mass, and the configuration evolves towards a
particlelike structure [5].
    % FIGURE 4
Its geometrical (gravitational) size can be estimated
from curvature considerations.
The scalar curvature $R$ is negative near $r\rightarrow 0$
and proportional to $r_0^{-2}$,
where $r_0$ is de Sitter
horizon defined by $\sqrt{3c^2/8\pi G\rho_{vac}}$.
Exterior curvature is positive, $R\sim{r_g r^{-3}}$,
where $r_g=2Gm/c^2$ and $m$ is the gravitational mass.
Then a surface of zero curvature, $r\sim{(r_0^2r_g)^{1/3}}$, can be chosen
as characteristic  size for a particlelike structure $r_p$.
In the particular model of Ref [5] it is determined by
$$r_p=(4r_0^2 r_g/3)^{1/3}=\frac{l_{Pl}}
{\pi^{1/3}}\biggl(\frac{m}{m_{Pl}}\biggr)^{1/3}
\biggl(\frac{{\rho}_{Pl}}{{\rho}_{vac}}\biggr)^{1/3}$$
\begin{figure}
\vspace{-4.0mm}
\begin{center}
\begin{tabular}{c c}
 \hspace{-3.0mm}
\rotatebox{00}{
 \epsfig{file=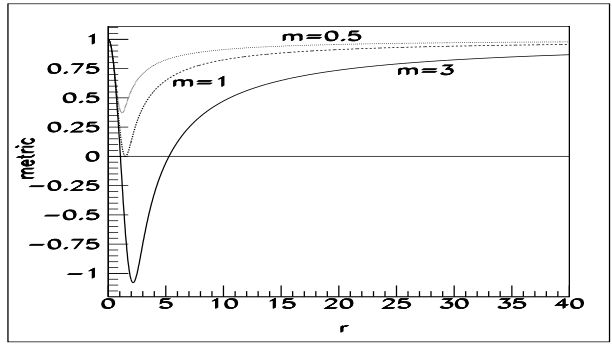 ,width=8.0cm,height=4.5cm}} & \\
% 
%\hspace{-6.0mm}
\end{tabular}
\end{center}
\caption{
De Sitter-Schwarzschild  configurations. The mass
is normalized to $m_{cr}$.}
\label{fig.1}
\end{figure}
In the context of spontaneous symmetry breaking
de Sitter core can be related to the vacuum expectation value
$v$ of a Higgs field giving mass to a particle.
Then $\rho_{vac}/\rho_{Pl}=\lambda v^4/4$,
where
$\lambda$ is self-coupling of a Higgs field.
It gives a particle mass $m=gv$, where $g$ is the relevant
coupling. The ratio of geometric size to a quantum size
${\lambda}_c$ (Compton
wave length) is given by
$$\frac{r_p}{{\lambda}_c}=\biggl(\frac{4g^4}{\pi\lambda}\biggr)^{1/3}$$
Taking into account that for Higgs scalar $g=\sqrt{2\lambda}$, we can
estimate its self-coupling $\lambda$ from the requirement that geometrical size of a
particle can not be bigger than its quantum size.
It gives  $\lambda\leq{\pi/16}$, which allows us to find the upper
limit for a mass of the Higgs scalar.
In the Weinberg-Salam theory $v=246$ GeV, and we get for the Higgs mass $m_H\leq{154}$ GeV.\\
With the upper bound for self-coupling $\lambda$,
we can estimate upper limits for sizes of leptons. It gives
$r_e<{1.91\times{10^{-18}}}$ cm,
$r_{\mu}<{1.13\times{10^{-17}}}$cm, and $r_{\tau}<{2.89\times{10^{-17}}}$cm.
To estimate a lower limit on a size of a particle we take into account
a possible cosmological scenario of particle production in the course
of phase transitions in the early Universe. In this context the limiting
scale for a vacuum expectation value $v$ is the scale at which Compton
wave length ${\lambda}_c$ of a particle fits within the causal horizon
$r_0=H^{-1}=\sqrt{3c^2/8\pi G\rho_{vac}}$. It gives the lower limits
for FP sizes as $r_e>6.21\times{10^{-26}}$cm, $r_{\mu}>1.05\times{10^{-26}}$cm,
and $r_{\tau}>4.10\times{10^{-27}}$cm.
{\bf References}
\newline
[1] ICHEP98, XXIX International Conference on High Energy Physics.
\newline   
[2] F. Abe et al Phys.Rev. {\bf D55} (1997) R5263
\newline
[3] Eur. Phys. J.{ \bf C3} (1998) 279
\newline
[4] K. Maeda, T. Tashizawa, T. Torii, M. Maki, Phys.Rev.Lett. {\bf 72} (1994) 450
\newline
[5] I. Dymnikova, Gen.Rel.Grav.{\bf 24} (1992) 235; Int.J.Mod.Phys. {\bf D5} (1996) 529
\end{document}